\documentclass[aps,pre,twocolumn,groupedaddress, showpacs]{revtex4-1}

\usepackage{graphicx}
\usepackage{dcolumn}
\usepackage{bm}
\usepackage[dvips]{color}

\def\mrm{\mathrm}
\def\mbf{\mathbf}

\def\goto{\rightarrow}

\def\qq{\mbf{q}}
\def\rr{\mbf{r}}

\def\mrm{\mathrm}
\def\mbf{\mathbf}
\def\goto{\rightarrow}

\def\qq{\mbf{q}}
\def\rr{\mbf{r}}
\def\VV{\mbf{V}}
\def\del{\partial}
\def\D{{\it \Delta}}

\def\deli0{\delta_{\sigma_i 0}}
\def\delj0{\delta_{\sigma_j 0}}

\def\b_ex{\beta_\mrm{ext}}

\definecolor{darkgreen}{rgb}{0.0, 0.33, 0.0}

\def\n+{_{n+1}}

\def\AA{\vec{A}}
\def\DAA{ {\it \Delta} \! \vec{A}}
\def\XX{\vec{X}}

\def\bb{\vec{b}}
\def\VV{\vec{V}}
\def\jj{\vec{j}}

\begin{document}


\preprint{APS/123-QED}

\title{
Usefulness of an equal-probability assumption 
for out-of-equilibrium states: \\
a master equation approach
}

\author{Tomoaki Nogawa}
\email{nogawa@serow.t.u-tokyo.ac.jp}
\affiliation{%
Department of Applied Physics, 
The University of Tokyo, Hongo, Bunkyo-ku, Tokyo 113-8656, Japan
}%

\author{Hiroshi Watanabe}
\affiliation{%
Institute for Solid State Physics, The University of Tokyo, Kashiwanoha 5-1-5, Kashiwa,
Chiba 277-8581, Japan
}%

\author{Nobuyasu Ito}
\affiliation{%
Department of Applied Physics, 
The University of Tokyo, Hongo, Bunkyo-ku, Tokyo 113-8656, Japan
}%

\begin{abstract}
We examine the effectiveness of assuming an equal probability 
for states far from equilibrium. 
For this aim, we propose a method to construct a master equation for extensive variables 
describing non-stationary nonequilibrium dynamics. 
The key point of the method is the assumption that 
transient states are equivalent to the equilibrium state 
that has the same extensive variables, i.e., 
an equal probability holds 
for microscopic states in nonequilibrium. 
We demonstrate an application of this method to the critical relaxation 
of the two-dimensional Potts model by Monte Carlo simulations. 
While the one-variable description, which is adequate for equilibrium, 
yields relaxation dynamics that are very fast, 
redundant two-variable description well reproduces 
the true dynamics quantitatively.
These results suggest that some class of the nonequilibrium state can be described 
with a small extension of degrees of freedom, 
which may lead to an alternative way to understand nonequilibrium phenomena.
\end{abstract}

\maketitle

\section{Introduction}

An surprising property of equilibrium states is that 
they can be identified by a small number of extensive variables (EVs) 
even though the system has an almost infinite number of degrees of freedom (DOFs). 
When EV values are given, 
the equilibrium system almost surely exhibits the most typical state, 
i.e., the equilibrium state is uniquely determined by macroscopic observation. 
This is the basis of the equilibrium statistical mechanics 
as the law of large numbers and an equal probability for microscopic states. 
On the other hand, nonequilibrium states generally need more information to be identified, 
and a history dependence (non-Markov process) appears if we label the state 
in a manner similar to that used to label the equilibrium states \cite{Mori65}. 
To treat the dynamics of a system as a Markov process, 
one has to deal with more DOFs, 
equal to the total number of the system in the worst case. 
However, it is natural to expect that there will be many situations 
in which a small extension of the DOFs will provide 
a sufficiently accurate description of nonequilibrium phenomena. 
In this paper, we propose a method to describe the non-stationary nonequilibrium dynamics 
by employing a few more EVs that are redundant for equilibrium states. 
We start from a macroscopic description with a small number of DOFs 
and approach the microscopic description by increasing degrees.
This approach proceeds in the manner opposite to the usual approach, 
which start from microscopic description and reduces the DOFs 
although approaching the same point. 
Thus, this approach may be a new method to connect the microscopic and macroscopic worlds 
and add a new perspective to nonequilibrium statistical physics.

If out-of-equilibrium dynamics can be described 
by the information of equilibrium states, 
like the successful linear response theory~\cite{Kubo91}, 
it can be very useful for our understanding. 
In the vicinity of the equilibrium, we have the phenomenological theory stating 
that the thermodynamic force that restores the system to an equilibrium 
is proportional to the free energy gradient, 
which is a function of EVs~\cite{Onsager31a, Onsager31b, Doi11}. 
A quadratic well of the free energy is often assumed, 
which leads to a linear response to the deviation from its lowest point, 
i.e., the equilibrium point. 
This theory is useful for a weak nonequilibrium, 
but cannot be applied for a strong nonequilibrium.

The projection operator method \cite{Nakajima58, Zwanzig60, Mori65} 
is typically used to extract the dynamics of slow-changing variables, 
i.e., integrating out the rapidly changing variables. 
In the formalism by Mori \cite{Mori65}, 
the projection leads to a generalized Langevin equation 
as an equation of motion for the remaining variables. 
The reduction of the number of DOFs can be seen as a local equilibrium approach. 
We take a similar approach; specifically, we develop a method to describe stochastic dynamics.

Consider the case of the relaxation dynamics far from equilibrium, 
where intensive variables such as temperature are controlled. 
It is difficult to relate a general transient state to the equilibrium state 
that corresponds to the external condition. 
However, a transient state labeled by a set of EVs, such as internal energy, 
can be related to another equilibrium state having the same energy. 
It is a microcanonical ensemble that is equivalent to the canonical ensemble 
at a temperature different from that of the environment. 
As mentioned before, nonequilibrium states need a higher number of EVs 
than equilibrium states. 
Here we call such states ``extended equilibrium states''. 
The equivalence of the transient state to the extended equilibrium state 
means that an equal probability holds for the nonequilibrium state. 
It is worth considering the possibility of describing nonequilibrium dynamics 
by using only information from the extended equilibrium state. 
If such description is possible, it will be a useful clue 
for understanding nonequilibrium phenomena 
based on the well-understood equilibrium property.

This paper is organized as follows. 
In the next section, we propose a method to construct a master equation 
by using only the equilibrium and macroscopic information. 
In Secs.~\ref{sec:mean_field}, we analyze the all-to-all coupling Ising model 
as the simplest example. 
In Sec.~\ref{sec:1var} and \ref{sec:2var}, 
we apply our approach to the three-state Potts model on the square lattice 
with one- and two-variable descriptions, respectively.
Section~\ref{sec:discussion} presents the discussion and perspective.

\section{Method}

\subsection{Master equation}

We consider a dynamics represented by a master equation 
in terms of a microscopic state variable 
$\XX = (\sigma_1, \sigma_2, \cdots, \sigma_N)$ as 
\begin{eqnarray}
\frac{\del p( \XX, t )}{\del t} = \sum_{\XX'} \left[
\tilde{w}(\bb; \XX | \XX' ) p( \XX', t )
- \tilde{w}(\bb; \XX' | \XX ) p( \XX, t ) \right], 
\label{eq:master0}
\end{eqnarray}
where $p(\XX,t)$ is the probability distribution function (PDF) of $\XX$ at time $t$, 
and $\tilde{w}(\bb; \XX' | \XX)$ is the transition probability per unit time 
from $\XX$ to $\XX'$ with given $\bb$. 
Here $\bb=(b_1, b_2, \cdots b_n)$ is a set of intensive variables,  
such as inverse temperature, magnetic field, or chemical potential, 
introduced as an external control parameter. 
Our aim is to derive the corresponding master equation 
in terms of the EVs $\AA=(A_1, A_2, \cdots, A_n)$ that are conjugate to $\bb$, 
such as internal energy, magnetization, or number of particles.
$\AA$ is a function of $\XX$, which we denote as $\AA = \AA_{\XX}$; 
thus, the PDF of $\AA$ is given by 
\begin{equation}
P(\AA, t) = \sum_{\XX} p(\XX, t) \delta_{\AA_{\XX} \AA} \ . 
\end{equation}

The main assumption of the present formulation is the equal probability 
for microstates for any given $\AA$ and $t$ as 
\begin{equation}
p(\XX,t) = P(\AA_{\XX},t)/g(\AA), 
\label{eq:equal_prob}
\end{equation}
where $g(\AA)$ is the number of states having $\AA$ 
as $g(\AA) \equiv \sum_{\XX} \delta_{\AA_{\XX} \AA }$. 
By multiplying Eq.~(\ref{eq:master0}) by $\delta_{\AA_{\XX} \AA}$ 
and taking the summation over $\XX$, 
the left hand side yields $\del P(\AA,t)/\del t$. 
This represents a projection from the $\XX$-space to the $\AA$-space. 
The second term on the right hand side becomes 
\begin{eqnarray}
& & \sum_{\XX, \XX'} \tilde{w}(\bb; \XX'|\XX) p(\XX,t) \delta_{\AA_{\XX} \AA}
\nonumber \\
& = & \sum_{\AA'} \sum_{\XX, \XX'} \delta_{\AA'_{\XX'} \AA'} 
\tilde{w}(\bb; \XX'|\XX) \delta_{\AA_{\XX} \AA} p(\XX,t)
\nonumber \\
& = & \sum_{\AA'} W(\bb; \AA'|\AA) P(\AA,t).
\end{eqnarray}
To obtain the last line, we use Eq.~(\ref{eq:equal_prob}) and put 
\begin{eqnarray}
W(\bb; \AA' | \AA ) &\equiv& \frac{1}{g(\AA)}  \sum_{\XX, \XX'} 
\delta_{\AA_{\XX'} \AA' }
\tilde{w}(\bb; \XX'| \XX)
\delta_{\AA_{\XX} \AA  }
\nonumber \\
&=& \left \langle \sum_{\XX'} 
\delta_{\AA_{\XX'} \AA' } \tilde{w}(\bb; \XX'| \XX) 
\right \rangle_{\!\! \AA} ,
\label{eq:W0}
\end{eqnarray}
where the bracket indicates the unweighted average with $\AA$ fixed as 
\begin{eqnarray}
\left \langle f(\XX) \right \rangle_{\!\!\AA} \equiv 
\frac{1}{g(\AA)} \sum_{\XX} f(\XX) \delta_{\AA_{\XX} \AA } \ . 
\end{eqnarray}
By performing the same operation to the first term in the right hand side 
of Eq.~(\ref{eq:master0}), we obtain the following form closed for $P(\AA,t)$ as 
\begin{eqnarray}
\frac{\del P( \AA, t )}{\del t} = \sum_{\AA'} \left[
W(\bb; \AA | \AA' ) P( \AA', t )
- W(\bb; \AA' | \AA ) P( \AA, t ) \right] .
\label{eq:master}
\end{eqnarray}

It is often the case that the transition probability $\tilde{w}$ 
is expressed as 
\begin{eqnarray}
\tilde{w}(\bb; \XX'| \XX) = w(\bb; \AA_{\XX'}-\AA_{\XX} ) D(\XX'|\XX), 
\label{eq:factorization}
\end{eqnarray} 
where the function $D(\XX'|\XX)$ equals unity if a direct (one-step) transition path exists 
between $\XX$ and $\XX'$, and is zero otherwise. 
For example, $w(\bb; \DAA)$ is proportional to $\mrm{min}[ 1, e^{-\bb \cdot \DAA} ]$ 
with $\DAA=\AA_{\XX'}-\AA_{\XX}$ in the Metropolis algorithm. 
The expression Eq.~(\ref{eq:factorization}) reflects the fact that 
the change of $\AA$ is decomposed to the local change by the short-range interaction, 
which is not explicitly dependent on the accumulated value $\AA$ itself.
In such a case, $W$ is expressed as 
\begin{eqnarray}
W(\bb; \AA+\DAA | \AA ) = \mu(\AA; \DAA ) w(\bb; \DAA) 
\label{eq:trans_prob}
\\
\mrm{with} \quad
\mu(\AA; \DAA ) = \left\langle \sum_{\XX'}
\delta_{\AA_{\XX'} \AA+\DAA } D(\XX'|\XX)
\right \rangle_{\!\!\AA}. 
\end{eqnarray} 
Here $\mu(\AA;\DAA)$ is regarded as the (un-normalized) PDF 
of the possible path, causing $\DAA$ for a given $\AA$. 
Note that $\mu$ itself is not a dynamical variable 
but a state quantity that is statistically evaluated 
in the extended equilibrium ensemble mentioned before. 
Thus, the information of the spatial configuration is reduced to a small number of quantities: 
the components of $\DAA$.

Since the general nonequilibrium state requires a macroscopic number of DOFs, 
the present master equation to resemble a Markov process with finite DOFs 
is an approximated one. 
However, it is possible to improve its accuracy in a systematic manner 
by increasing the number of EVs.
In our strategy, $\mu$ is estimated from the equilibrium ensemble: 
an equal probability with fixed EVs $\AA$. 
Although the factorization of Eq.~(\ref{eq:trans_prob}) is practically very useful 
and naturally derived for standard Monte-Carlo (MC) dynamics, 
it is not clear whether it can be used for any system. 
If it cannot be used, we have to estimate $W(\bb; \AA+\DAA | \AA )$ 
for every pair of $\bb$ and $\AA$ as in Eq.~(\ref{eq:W0}).

One may think that the dynamics with the transition probability of 
Eq.~(\ref{eq:trans_prob}) with $w(\bb; \DAA) \propto \mrm{min}[ 1, e^{-\bb \cdot \DAA} ]$ 
is similar to that with 
\begin{equation}
W(\bb; \AA' | \AA) \propto \mrm{min}[ 1, e^{-[F(\bb; \AA') - F(\bb; \AA)]} ]
\label{eq:trans_prob2}
\end{equation}
where $F(\bb; \AA) \equiv \bb \cdot \AA - \ln g(\AA)$ 
is the dimensionless free energy, and $g(\AA)$ is the number of states having $\AA$. 
Such an approach was examined in Refs.~\cite{Lee95, Shteto97, Shteto99}. 
Although both of Eqs.~(\ref{eq:trans_prob}) and Eq.~(\ref{eq:trans_prob2}) 
lead to the equilibrium PDF $P_\mrm{eq} \propto e^{-F( \bb; \AA)}$ 
as a stationary distribution, the time evolutions are different. 
Our strategy can construct the equation of motion for the EVs 
from first-principles, as shown below.
The resultant equation depends on the function form of $w(\bb, \DAA)$ 
and does not includes a fitting parameter, such as a dissipation coefficient.



\subsection{Equation of motion for the EVs}

Let $n$ be the dimension of $\bb$ and $\AA$. 
At equilibrium, $\AA$ is a unique function of $\bb$ except for the multistable points. 
[On the other hand, $\bb$ is a unique function of $\AA$ even in coexisting phases.] 
The equilibrium state of a given $\bb$ is represented by a point 
in the $n$-dimensional space, which we call the $A$-space.
On the other hand, the trajectory of nonequilibrium dynamics 
is given by a curved line. 
Generally, each point of the trajectory, i.e., a transient state, 
is not related to the equilibrium state with $\bb$ of the environment. 

We can define the velocity field $\VV=(V_{A_1}, \cdots, V_{A_n})$ as 
\begin{equation}
\VV( \bb; \AA ) \equiv 
\sum_{\DAA} \DAA \, W(\bb; \AA + \DAA), 
\end{equation}
which is used to express the probability current as 
$\jj(\bb; \AA, t) = P(\AA,t) \VV(\bb; \AA)$. 
Note that $\VV$ is independent of time.
The dynamics is deterministic in the thermodynamic limit, 
where the fluctuation of the EVs can be ignored in comparison with the mean values.
The time evolution of 
$\langle \AA \rangle(t) \equiv \int \mrm{d}\AA \AA P(\AA,t)$ 
can then be given as a solution of the equation of motion, 
\begin{equation}
\frac{\mrm{d} \langle \AA \rangle}{\mrm{d} t} = \VV( \bb; \langle \AA \rangle ).
\label{eq:eom}
\end{equation}
Hereafter, we only treat such most probable dynamics.

If the rotation $(\del/\del \AA) \times \VV$ equals zero, 
we can define a potential function $F_\mrm{neq}(\bb; \AA)$, 
which can be regarded as nonequilibrium free energy, as 
\begin{equation}
\VV(\bb; \AA) = - \frac{ \del }{ \del \AA} F_\mrm{neq}(\bb; \AA), 
\label{eq:F_neq}
\end{equation}
which depends on the details of the dynamics represented by $w(\bb; \DAA)$. 
This is in contrast to the dynamics corresponding to Eq.~(\ref{eq:trans_prob2}), 
which is governed by the equilibrium free energy.

\subsection{Equilibrium ensemble with fixed EVs}

To construct the master equation, 
we need to calculate $\mu$ for various $\AA$ values 
as a statistical distribution in the extended equilibrium ensemble.
We calculate it by MC simulations.
It is not efficient to perform an independent simulation for each $\AA$. 
Instead, we perform the Wang--Landau sampling \cite{Wang01, Landau04} 
to obtain a flat histogram of $\AA$.
The Wang--Landau sampling enables us to cover a wide range of $\AA$ uniformly by a single run.
In addition, a joint density (number) of states $g(\AA)$ is obtained as a byproduct, 
which leads to the {\it microcanonical} entropy $\ln g(\AA)$. 
The generalized dimensionless free energy is given by 
$F(\bb; \AA) = \bb \cdot \AA - \ln g(\AA)$, and 
its extremal condition, $\del F / \del \AA=0$, gives an equilibrium relation, 
$\bb = \del \ln g(\AA) / \del \AA$ \cite{Nogawa11b}.

\section{The all-to-all coupling Ising model}
\label{sec:mean_field}

Here we demonstrate how the above method is applied 
to an analytically solvable model. 
Consider a two-state Potts model, i.e., the Ising model 
containing $N$ spins with all-to-all coupling. 
The energy function is 
\begin{eqnarray}
E = -\frac{1}{N} \sum_{i,j} \delta_{\sigma_i \sigma_j} 
= -\frac{1}{N} [ N_0^2 + ( N - N_0 )^2 ].
\end{eqnarray}
The spin variables $\{ \sigma_i \}$ take the value 0 or 1.
The number of spins at state $0$ is defined as
\begin{eqnarray}
N_0 = \sum_i \delta_{ \sigma_i 0 } \equiv N n_0.
\end{eqnarray}
Since the present system does not have a meaningful spatial structure, 
the state can be accurately identified only by $N_0$. 
From the $Z_2$-symmetry, it is sufficient to consider $N_0 \ge N/2$. 
Therefore, the magnetization and energy cannot change independently 
but have a one-to-one relation. 
unlike infinite-dimensional systems. 
By utilizing this property, we use the notation as if 
$N_0$ were conjugate to inverse temperature $\beta$ in the following; 
$E$ and $\D E$ as the arguments of $w$ and $\mu$ are replaced 
with $N_0$ and $\D N_0$, respectively.

The number of states $g(N_0)$ is a function of $N_0$ as
\begin{eqnarray}
g(N_0) = \frac{N!}{N_0!(N-N_0)!},
\end{eqnarray}
and the dimensionless free energy per site is written as 
\begin{eqnarray}
\frac{F}{N}  & \approx & -\beta [ n_0^2 + (1-n_0)^2 ] 
\nonumber \\ && 
- \left[ \ln \frac{1}{1-n_0} - n_0 \ln \frac{n_0}{1-n_0} \right]
\\
& = & \ln 2 - \frac{\beta}{2} + \frac{1}{2} (1-\beta) m^2 + \frac{1}{6} m^4 + O(m^6),
\end{eqnarray}
where $\beta = 1/k_\mrm{B} T$ is an inverse temperature 
and $\ m \equiv 2 n_0 - 1$ is the order parameter.
It is well known that this is equivalent to the $\Phi^4$ free energy, 
i.e., the Landau model, which has a single minimum 
at $m=0$ for $\beta \le \beta_c = 1$ 
and has two symmetric minima at $\pm m(\beta)$ with $0<m(\beta)\le1$ 
for $\beta>\beta_c$;
$m(\beta) = \pm \sqrt{3(\beta-1)}$ for $m \ll 1$.

We consider the Metropolis dynamics; randomly picking up one spin 
and changing the state to 1 if it is 0 and vice versa. 
Therefore, the updated PDF is 
\begin{eqnarray}
\mu(N_0;+1) = 1 - n_0  \quad \mrm{and} \quad \mu(N_0;-1) = n_0.
\end{eqnarray}
The spin flip is accepted with probability rate 
\begin{equation}
\tilde{w}(\bb; \DAA) = \frac{N}{q-1} \min[1, e^{-\bb \cdot \DAA}], 
\label{eq:Metropolis_Potts}
\end{equation}
where $q=2$ for the Ising model. 
For $N_0 > N/2$, 
\begin{eqnarray}
w(\beta;+1) = N, \ 
w(\beta;-1) = N e^{-\beta \D E} 
\\
\mrm{with} \quad \D E = 2( 2 n_0 - 1 ) + O(N^{-1}). 
\end{eqnarray}
The velocity can now be written as 
\begin{eqnarray}
v_{N_0} &\equiv& V_{N_0}/N \nonumber \\
&=& \mu(N_0;+1) w(\beta;+1) - \mu(N_0;-1) w(\beta;-1)
\nonumber \\
&=& 1 - n_0[ 1 + e^{-2\beta(2 n_0 -1 )} ].
\end{eqnarray}
The time evolution is formally solved as 
\begin{eqnarray}
t = N \int_{n_0(0)}^{n_0(t)} \frac{dn_0}{v_{N_0}(\beta, n_0)}.
\label{eq:mf-master}
\end{eqnarray}
At the critical point $\beta=1$, we have $v_{N_0} = -m^3 + O(m^5)$, 
which leads to $m \propto t^{-1/2}$ for $m\ll 1$. 
The time evolution of energy per site behaves as $(E_c - E)/N \approx m^2/2 \propto t^{-1}$, 
where $E_c = N/2$ is the equilibrium energy at $\beta_c$.

\begin{figure}[t]
\begin{center}
\includegraphics[trim=20 40 140 20,scale=0.32,clip]{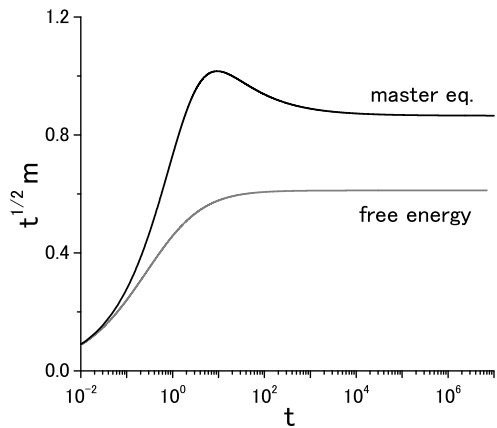}
\end{center}
\vspace{-5mm}
\caption{\label{fig:MF_E-t}
Time evolutions of order parameter $m=2n_0-1$ (multiplied by $t^{1/2}$) 
in the all-to-all coupling Ising model. 
The decimal logarithm is shown on the horizontal axis.
The result of the exact solution is denoted by ``master eq.'' and 
that obtained form the free-energy gradient is denoted by ``free energy''.
While the long-time behaviors of the two results are same as $m \propto t^{-1/2}$,
the short-time dynamics are significantly different. 
}
\end{figure}

We now consider the relation between the above dynamics and 
the free-energy landscape. 
The free-energy gradient is written as 
\begin{eqnarray}
-\frac{ \del F}{\del N_0} = 2 \beta( 2 n_0 - 1) - \ln \frac{n_0}{1-n_0}.
\label{eq:mf-freeenergy}
\end{eqnarray}
We have $-\del F/\del N_0 = -4m^3/3 + O(m^5)$ at $\beta=1$. 
Thus, the phenomenological equation of motion, $d N_0/dt = -\del F/\del N_0$, 
also yields $m \propto t^{-1/2}$. 
The short time dynamics is, however, distinctly different 
from the exact solution, Eq.~(\ref{eq:mf-master}). 
Especially, $\del F/\del N_0$ diverges for $n_0 = 0$ and $1$, 
as against the exact solution.
Figure~\ref{fig:MF_E-t} shows the results of the numerical integrations
of Eqs.~(\ref{eq:mf-master}) and (\ref{eq:mf-freeenergy}) 
with an initial condition $n_0 = 0.95$ at $t=0$. 
It is observed that the approximated dynamics with the equilibrium free energy 
is valid only in the vicinity of the destination state, i.e., the equilibrium.

\section{One-variable description}
\label{sec:1var}

Next, we investigate a finite-dimensional system: 
the ferromagnetic 3-state Potts model on the square lattice, 
which exhibits a second-order transition at the critical point
$\beta_c = \ln ( 1 + \sqrt{3} ) \approx 1.00505$. 
Here we did not choose the Ising model ($q=2$) 
to avoid an unusual critical behavior, 
the logarithmic divergence of specific heat.

We start with the minimal case, a description with one EV. 
For a system with fixed temperature, the most relevant EV 
is internal energy $E$, which is an EV conjugate to temperature. 
If we do not use it, the transition rate, 
which is a function of $\beta E$, cannot be determined. 
The energy $E$ is defined as 
\begin{eqnarray}
E = 2N - \sum_{\langle i,j \rangle \in \mrm{n.n.}} \delta_{\sigma_i \sigma_j}, 
\end{eqnarray}
where the summation is taken over all the nearest-neighbor pairs. 
Each spin ${\sigma_i}$ takes a value: 0, 1, or 2. 
The energy change of a single spin flip is in the range of $-4 \le \D E \le 4$. 
We performed the Wang--Landau sampling simulations \cite{Landau04} 
for samples with size $N = L \times L$ and $L = 64 - 1024$ 
by imposing a periodic boundary condition. 
We averaged the PDF $\mu(E;\D E)$ over four independent runs. 
In the estimation of $\mu$, we eliminated the update that the spin state remains the same. 
We also performed corresponding kinetic Monte Carlo (KMC) simulations 
\cite{Glauber63, Stoll73} as a non-approximated dynamics 
and compared the results with those obtained by the master equation. 
One MC step involves the following process $N$ times: 
picking up a spin randomly and flipping it to one of the $q-1$ states 
with probability $\mrm{min}[1, e^{-\bb \cdot \DAA}]$. 
Thus, the transition rate of $\DAA \ne \vec{0}$ is written as Eq.~(\ref{eq:Metropolis_Potts}).

\begin{figure}[t]
\begin{center}
\includegraphics[trim=20 40 140 20,scale=0.32,clip]{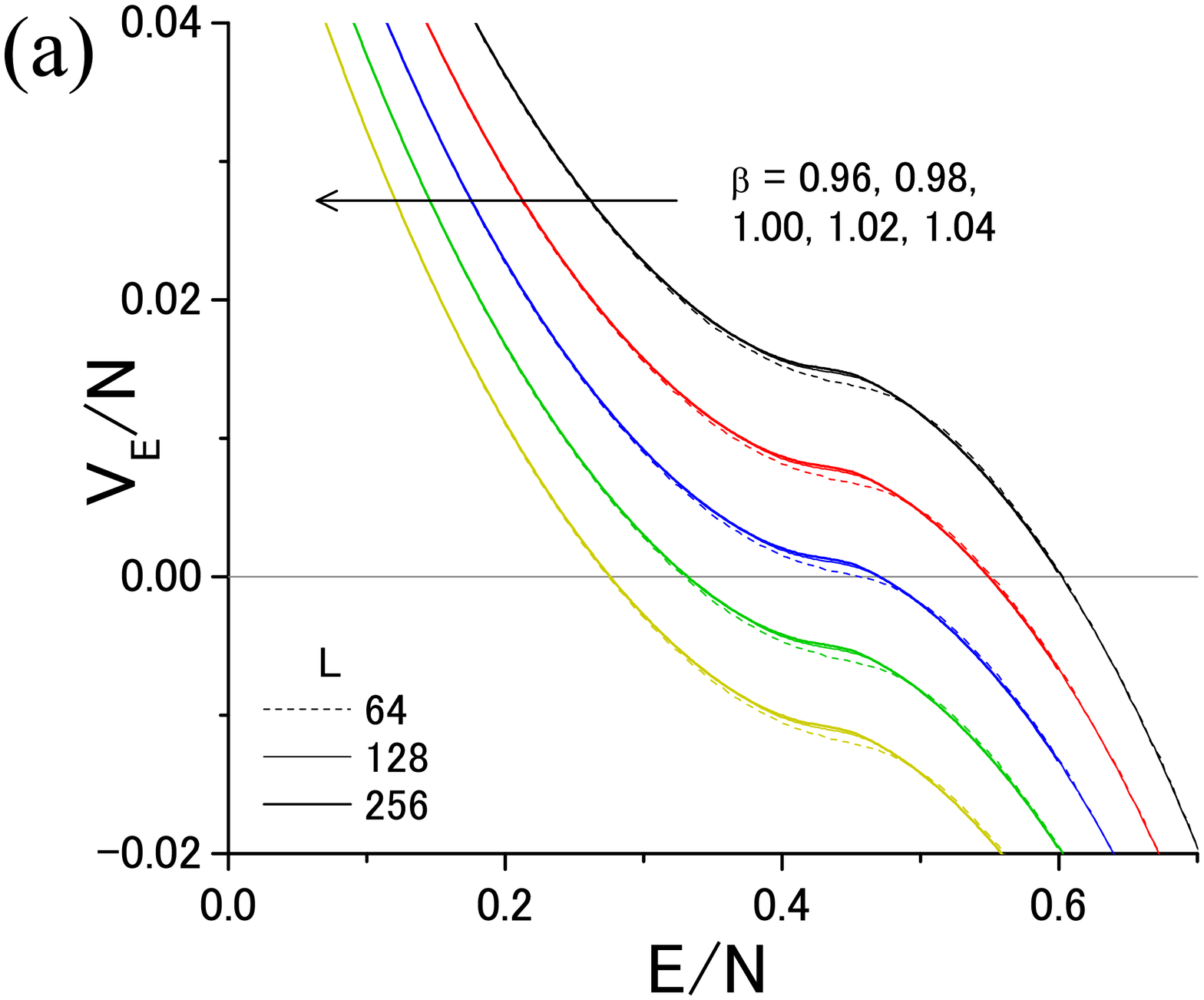}
\includegraphics[trim=20 40 140 20,scale=0.32,clip]{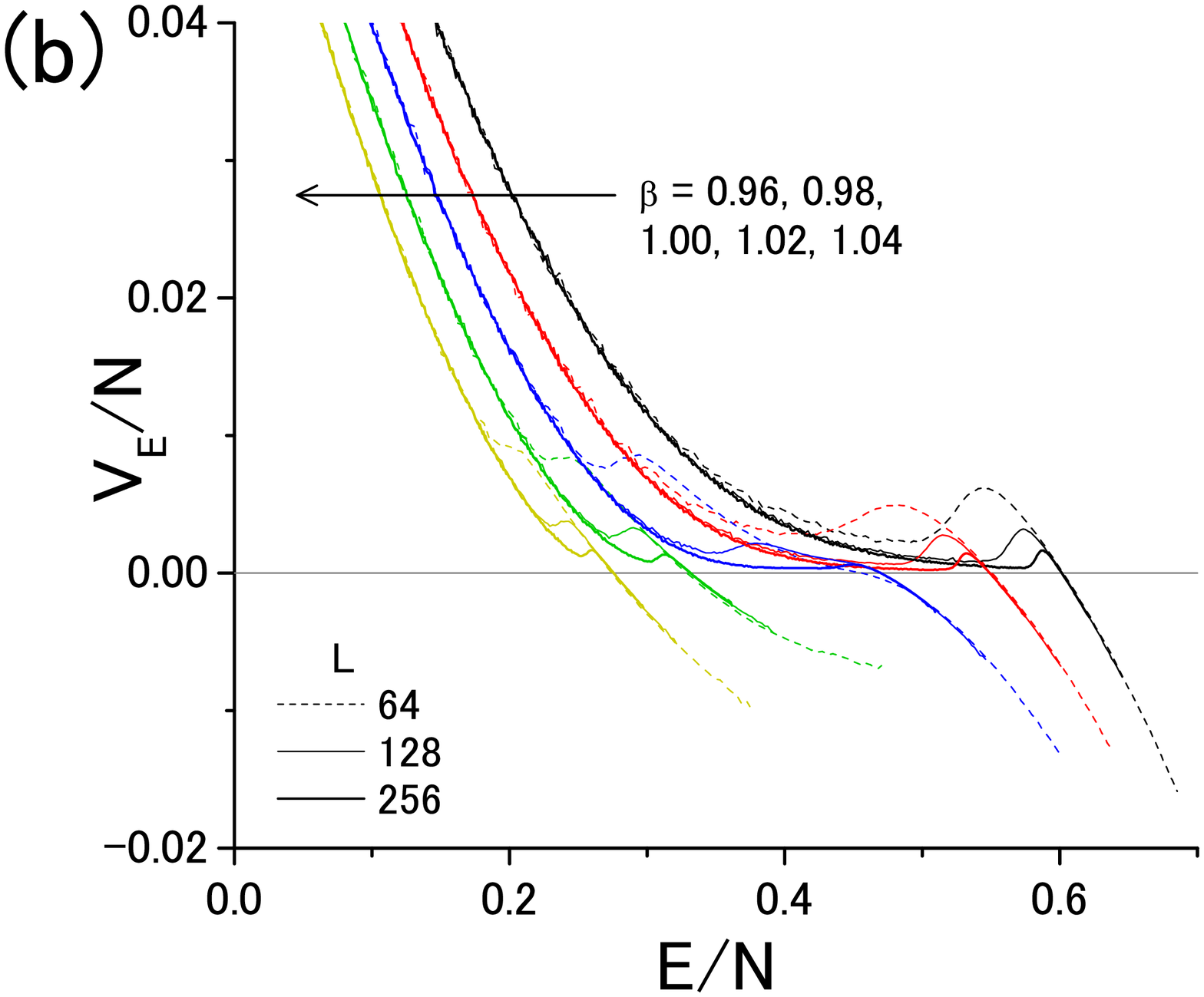}
\end{center}
\vspace{-5mm}
\caption{\label{fig:1d_V-E}
(color online) 
(a) Energy dependence of velocity for the one-variable description.
Colors denote temperatures.
(b) Velocity observed in the KMC. 
Colors denote temperatures.
Non-monotonic behaviors are observed and finite-size dependence 
appears near the zero-crossing points of $V_E$.
}
\end{figure}

\begin{figure}[t]
\begin{center}
\hspace{-6.8cm} {\bf{\large (a)}}\\ \vspace{-14.1pt}
\includegraphics[trim=20 40 140 20,scale=0.32,clip]{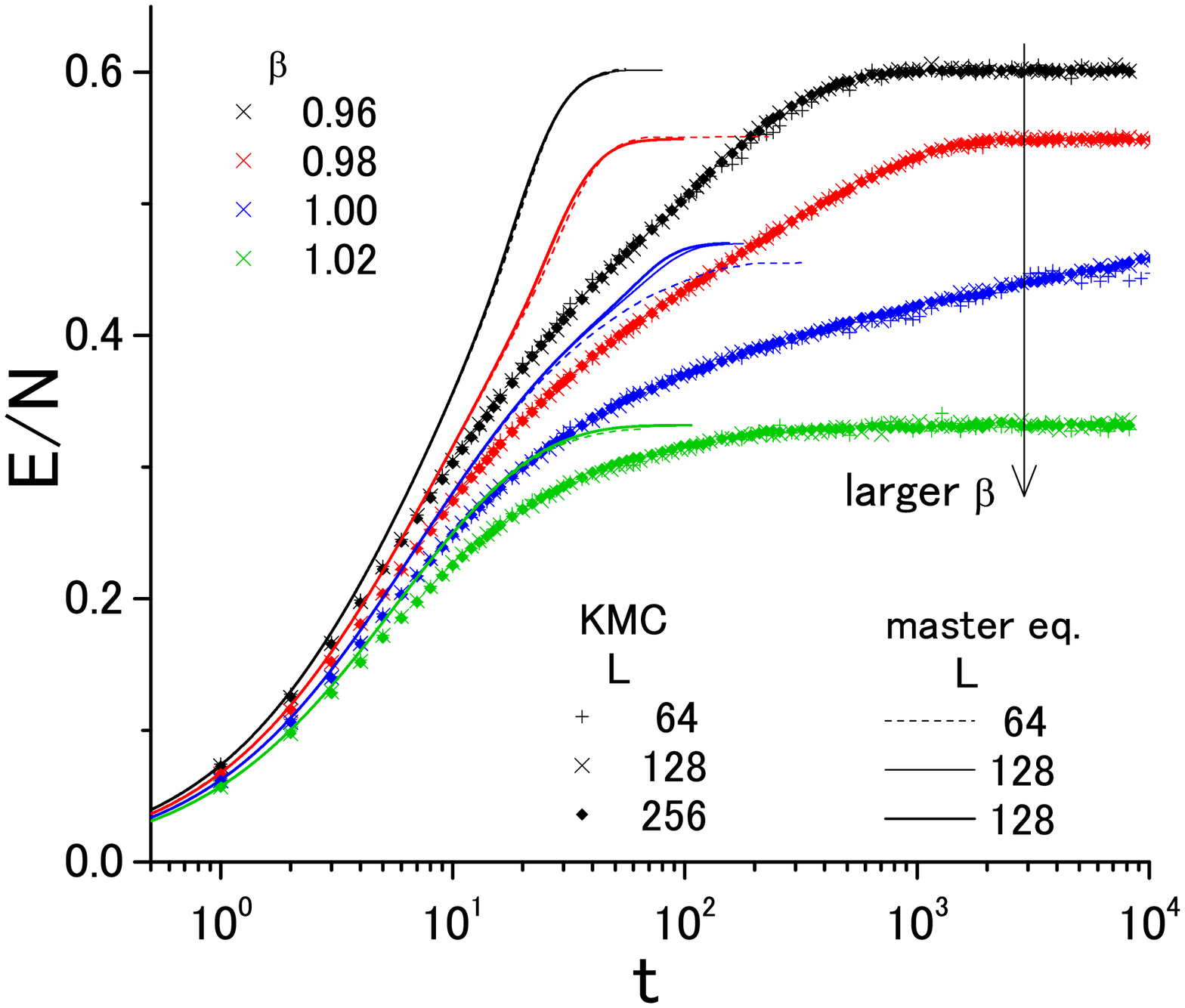}
\\ \hspace{-6.8cm} {\bf{\large (b)}}\\ \vspace{-14.1pt}
\includegraphics[trim=20 40 140 20,scale=0.32,clip]{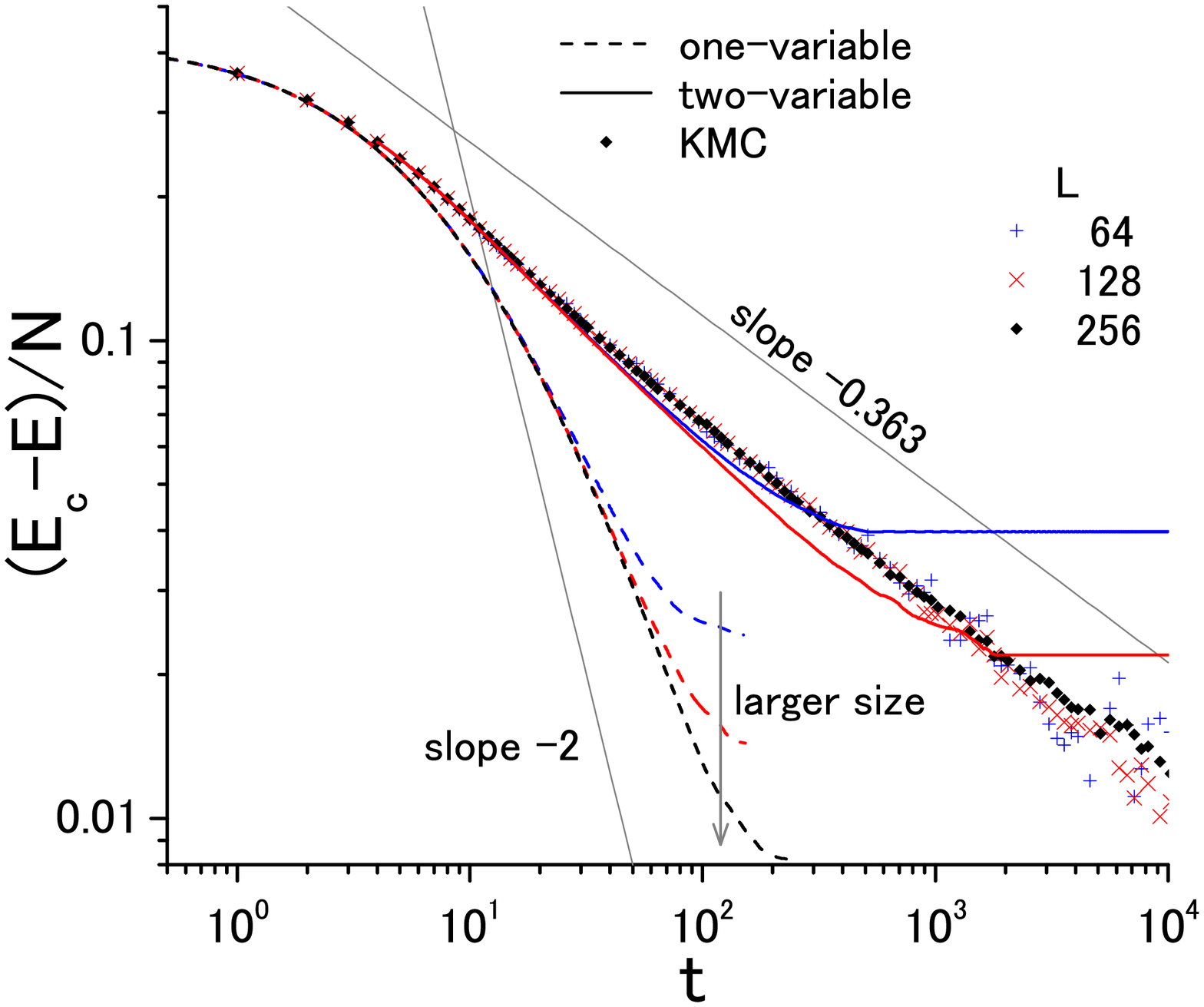}
\end{center}
\vspace{-5mm}
\caption{\label{fig:1d_E-t}
(color online) 
(a) Time evolution of energy.
Colors denote temperatures.
The decimal logarithm is shown on the horizontal axis.
The results of the one-variable description are denoted by lines
and those of KMC simulations are denoted by symbols.
While the equilibrium values are the same, the short-time dynamics
are significantly different between the two methods. 
(b) Relaxation of energy at the critical temperature.
The decimal logarithm is shown on the both axes. 
Colors denote temperatures.
Three groups are shown: the KMC (symbol), the one-variable description (dashed line), 
and the two-variable description (solid line).
}
\end{figure}

Figure~\ref{fig:1d_V-E}(a) shows the velocities $V_E(E)$ at several temperatures, 
which were obtained by the master equation with the one-variable
description.
Each zero-crossing point of $V_E$ is related to the equilibrium state at each temperature.
From these velocities, the time evolutions of the energies are obtained 
by the numerical integration like Eq.~(\ref{eq:mf-master}) 
and are plotted in Fig.~\ref{fig:1d_E-t}(a) 
along with the results of the KMC simulations. 
While the results obtained by the two methods have similar the equilibrium energy,
the relaxations obtained by the master equation are significantly faster 
than those obtained by the KMC method.

Figure~\ref{fig:1d_V-E}(b) shows the velocities calculated by the PDF $\mu$ 
obtained by nonequilibrium sampling through KMC simulations. 
These velocities reproduce the time evolution of the energy obtained by the KMC method. 
Non-monotonic behaviors observed just before $V_E$ crosses zero level arises 
as a result of the crossover from nonequilibrium sampling 
to equilibrium sampling in the long time regime, 
and the latter sampling coincides with the equilibrium velocity in Fig.~\ref{fig:1d_V-E}(a). 
When the system size is increased, this equilibrium region becomes smaller 
because the equilibrium fluctuation of $E/N$ disappears for $N \goto \infty$. 
In the thermodynamic limit, the velocities of equilibrium and nonequilibrium sampling 
are different for all $E$ except for the equilibrium point.
This means that the typical spin configuration of a given $E$, 
which is represented by $\mu$, is different between the two states.  
Thus, a one-variable is insufficient to describe this type of dynamics 
although it yields the correct equilibrium state.

\begin{figure}[t]
\begin{center}
\includegraphics[trim=20 40 140 20,scale=0.30,clip]{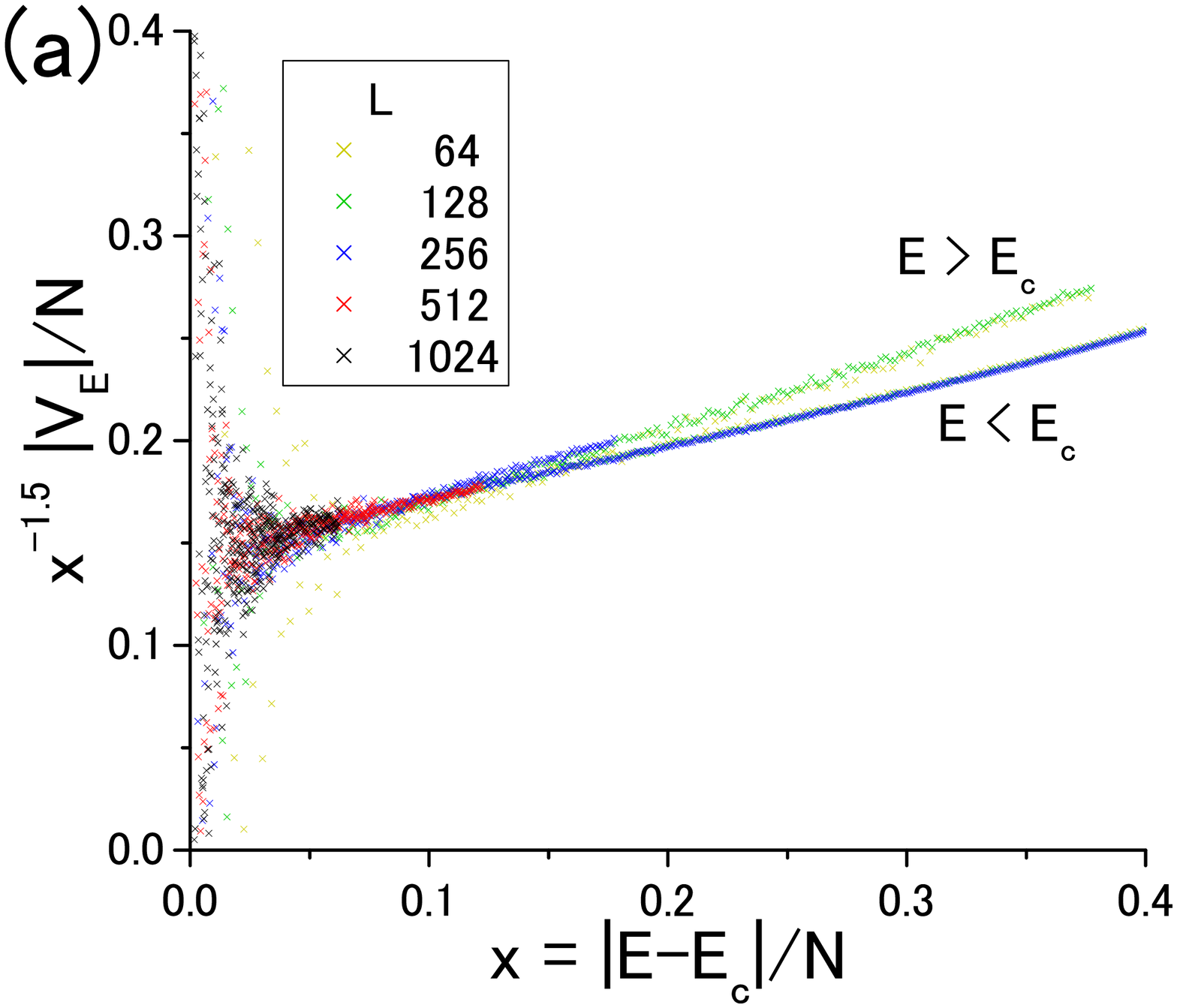}
\includegraphics[trim=20 40 140 20,scale=0.30,clip]{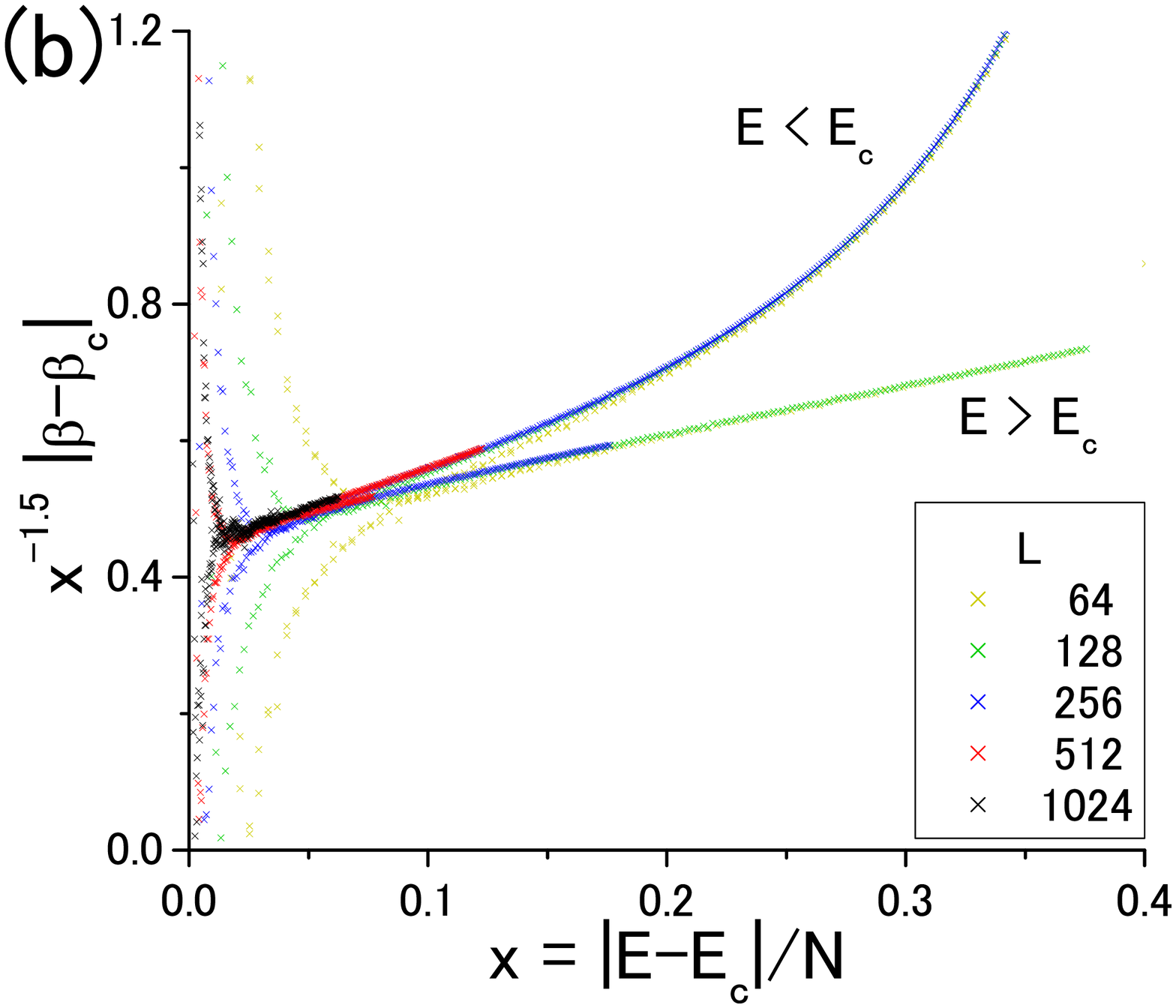}
\end{center}
\vspace{-5mm}
\caption{\label{fig:1d_V-Eb}
(color online) 
Confirmation of the power-law at $\beta=\beta_c$.
The microcanonical velocity (a) 
and microcanonical temperature (b). 
It is shown that $|V_E|$ and $|\beta-\beta_c|$ converge to finite values 
by multiplying $x^{1.5}$, where $x = |E-E_c|/N$, 
and taking the limit $x \goto 0$. 
This means the two quantities are proportional to $|E-E_c|^{1.5}$.
}
\end{figure}

As mentioned above, the dynamics obtained by the one-variable description fail 
even in the vicinity of the equilibrium point. 
Here we look at it more closely. 
Figure~\ref{fig:1d_V-Eb}(a) shows the relation between $V_E$ and $E$ 
at $\beta_c$, which indicates a power-law, $|V_E| \propto |E_c-E|^{1.5}$, 
in the limit: $E \goto E_c$. 
This leads to $E_c - E \propto t^{-2.0}$ 
although it is not confirmed in the short-time relaxation 
as shown in Fig.~\ref{fig:1d_E-t}(b) [dashed lines]. 
The exponent 2.0 is distinctly different from that obtained by the KMC method, 
$0.367$ \cite{Murase08, Nam08}.

Next, we examine the relation between the velocity and the free-energy gradient, 
$-\del F(\beta_c;E)/\del E = \beta(E) - \beta_c$. 
The microcanonical temperature $\beta(E)$ is defined as $\del \ln g(E)/\del E$. 
Figure~\ref{fig:1d_V-Eb}(b) shows a power law singularity 
$\beta(E) - \beta_c \propto ( E - E_c )^{1/(1-\alpha)}$, 
where $\alpha$ is the critical exponent of the specific heat. 
The exponent is consistent with the known value $3/2$ ($\alpha=1/3$). 
Therefore, the phenomenological dynamics using free energy also 
fail to reproduce the true dynamics in a similar manner to that using $V_E$. 
We can write the relation $V_E(E) \propto - \del F(\beta_c;E)/\del E$ 
since both the velocity and the gradient are proportional to $( E - E_c )^{1.5}$. 
Therefore, the nonequilibrium free energy defined in Eq.~(\ref{eq:F_neq}) 
is essentially equivalent to the equilibrium energy in the one-variable description. 
This will be discussed  more closely after seeing the two-variable description 
in the next section.


\begin{figure}[t]
\begin{center}
\hspace{-6.8cm} {\bf{\large (a)}}\\ \vspace{-14.1pt}
\includegraphics[trim=20 40 140 20,scale=0.32,clip]{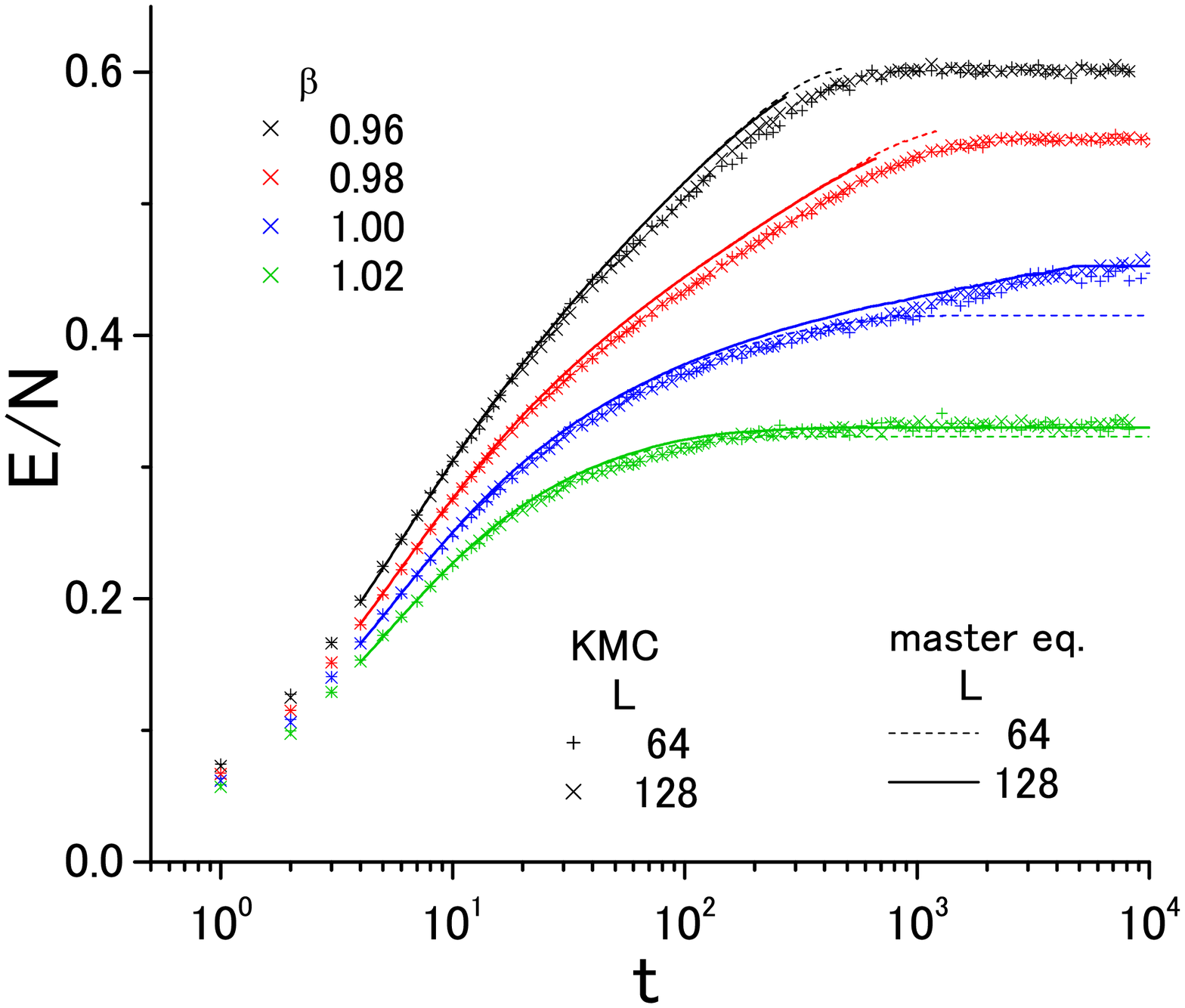}
\\ \hspace{-6.8cm} {\bf{\large (b)}}\\ \vspace{-14.1pt}
\includegraphics[trim=20 40 140 20,scale=0.32,clip]{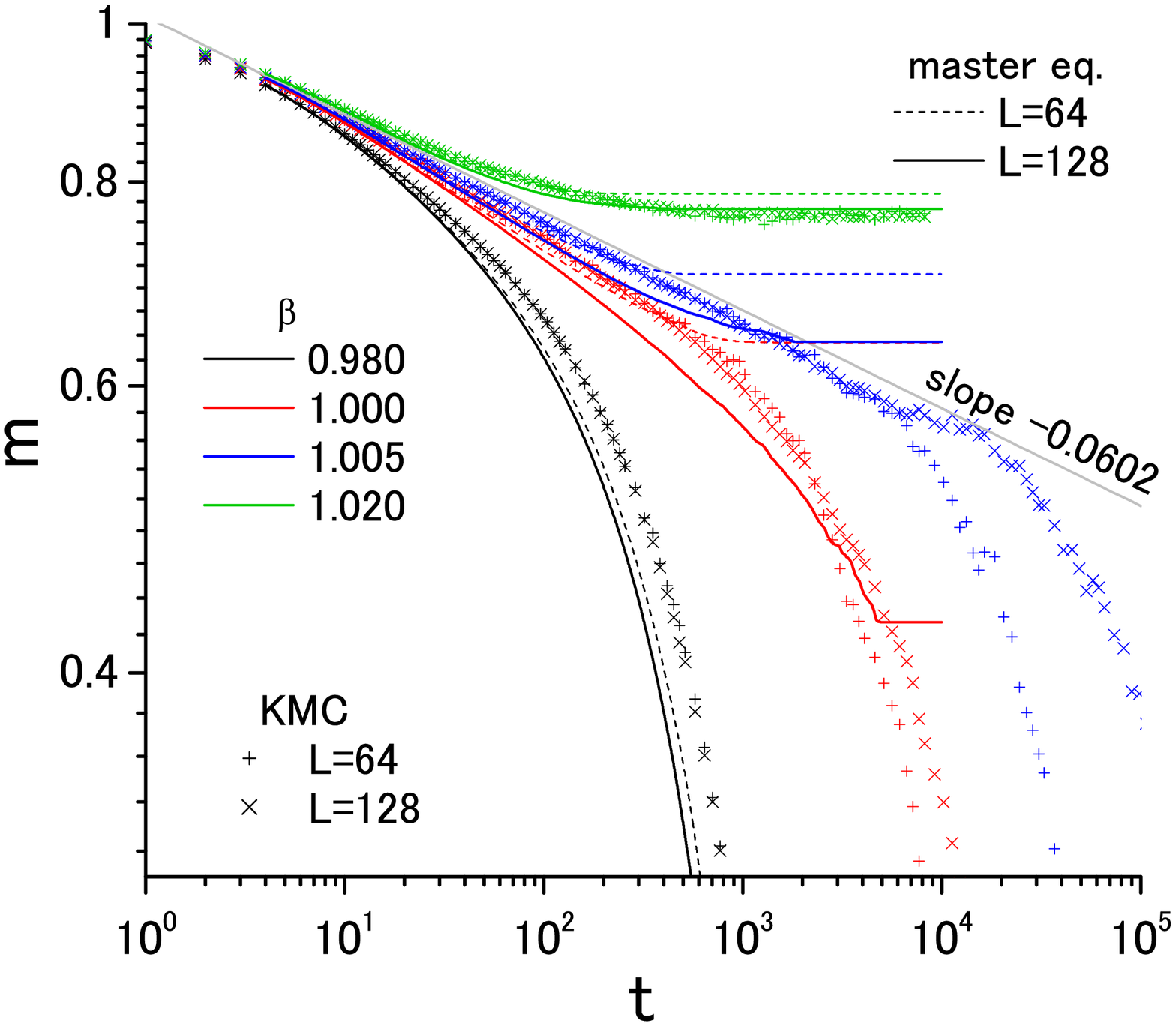}
\end{center}
\vspace{-5mm}
\caption{\label{fig:2d_E-t}
(color online) 
(a) Time evolution of the internal energy.
The results using the two-variable description (lines) 
and those obtained by the KMC method (symbols) are in good agreement. 
(b) Time evolution of the order parameter. 
The line with the slope $\beta/z \nu = 0.0602$ \cite{Nam08} is also shown.
The two-variable description reproduces the critical behavior well. 
}
\end{figure}

\section{Two-variable description}
\label{sec:2var}

Next, we employ one more EV, $N_0$, 
that is the number of spins taking the 0-state, 
which is related to the order parameter $m = (N_0/N - 1/q)/(1-1/q)$. 
The extensive and intensive variables are given 
by $\AA = (E, N_0)$ and $\bb=(\beta, -h)$, respectively. 
Here $h$ is the magnetic field coupled with the 0-state as $-h N_0$. 
In the following, we set $h=0$. 
The update of $N_0$ with a single spin flip is restricted to $-1 \le \D N_0 \le 1$.
For the two dimensional $A$-space, 
we calculate $\mu(E,M;\D E, \D N_0)$ whose total number of components
is proportional to  $N^2 \times 9 \times 3$. 
This requires such a large memory space during the Wang--Landau sampling 
that we divided $E$ and $N_0$ into bins of width $\sqrt{N}/4$. 
Due to this coarsegraining, 
the integration with the initial condition $E=0$ and $N_0=N$ was difficult 
because the trajectory runs near the upper edge of the $A$-space, 
$N_0 = 1 - E/4$, above which there is no state. 
As an initial condition, we used the values of $E$ and $N_0$ obtained by the KMC simulations
at $t=4$ for each temperature. 

We integrated the equation of motion, Eq.~(\ref{eq:eom}), 
with the 2nd order Runge-Kutta method with a discrete time step of 0.01. 
Because we had $\VV$ values only on the lattice points, 
we interpolated the value at the off-lattice points as 
$\VV(E,N_0) = \vec{c}_0 + \vec{c}_1 E + \vec{c}_2 N_0$, 
where the coefficients were determined by the three nearest lattice points.

The time evolution of $E$ is shown in Fig.~\ref{fig:2d_E-t}(a). 
The results of two different system sizes $L=64$ and $128$ 
are shown as dashed and solid lines, respectively.
In a regime where the two lines are identical, 
the behaviors can be considered to be in the thermodynamic limit.
The results of the two-variable description show a reasonably better agreement 
with those of the KMC method than the one-variable description result do  
(the relaxations are still slightly faster than those of the KMC).
The critical behavior is shown in Fig~\ref{fig:1d_E-t}(b) as solid lines. 
We found that the relaxations obtained by the master equation 
agree reasonably well with those of the KMC method.
The time evolutions of the order parameter are shown in Fig.~\ref{fig:2d_E-t}(b).
The relaxations obtained by the master equation and KMC are in good agreement 
as is the internal energy.

\subsection{Trajectory}

\begin{figure}[t]
\begin{center}
\includegraphics[trim=20 40 140 20,scale=0.32,clip]{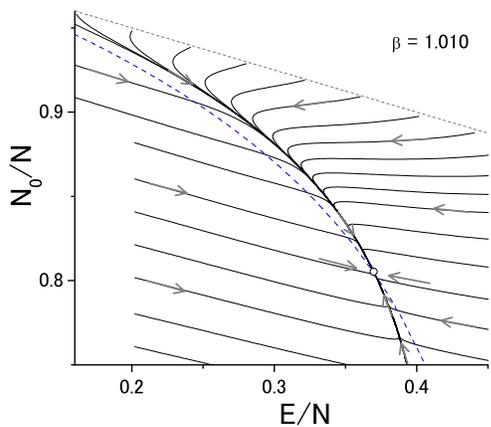}
\end{center}
\vspace{-5mm}
\caption{\label{fig:xy}
(color online) 
The trajectory in $(E \times N_0)$-space 
of $\b_ex=1.010 > \beta_c$ with various initial conditions. 
The fixed point is indicated by the open circle 
at $(E,N_0)/N = (0.370, 0.805)$. 
The dashed line denotes $h(E,N_0) = 0$. 
}
\end{figure}

\begin{figure}[t]
\begin{center}
\hspace{-6.8cm} {\bf{\large (a)}}\\ \vspace{-14.1pt}
\includegraphics[trim=20 40 140 20,scale=0.32,clip]{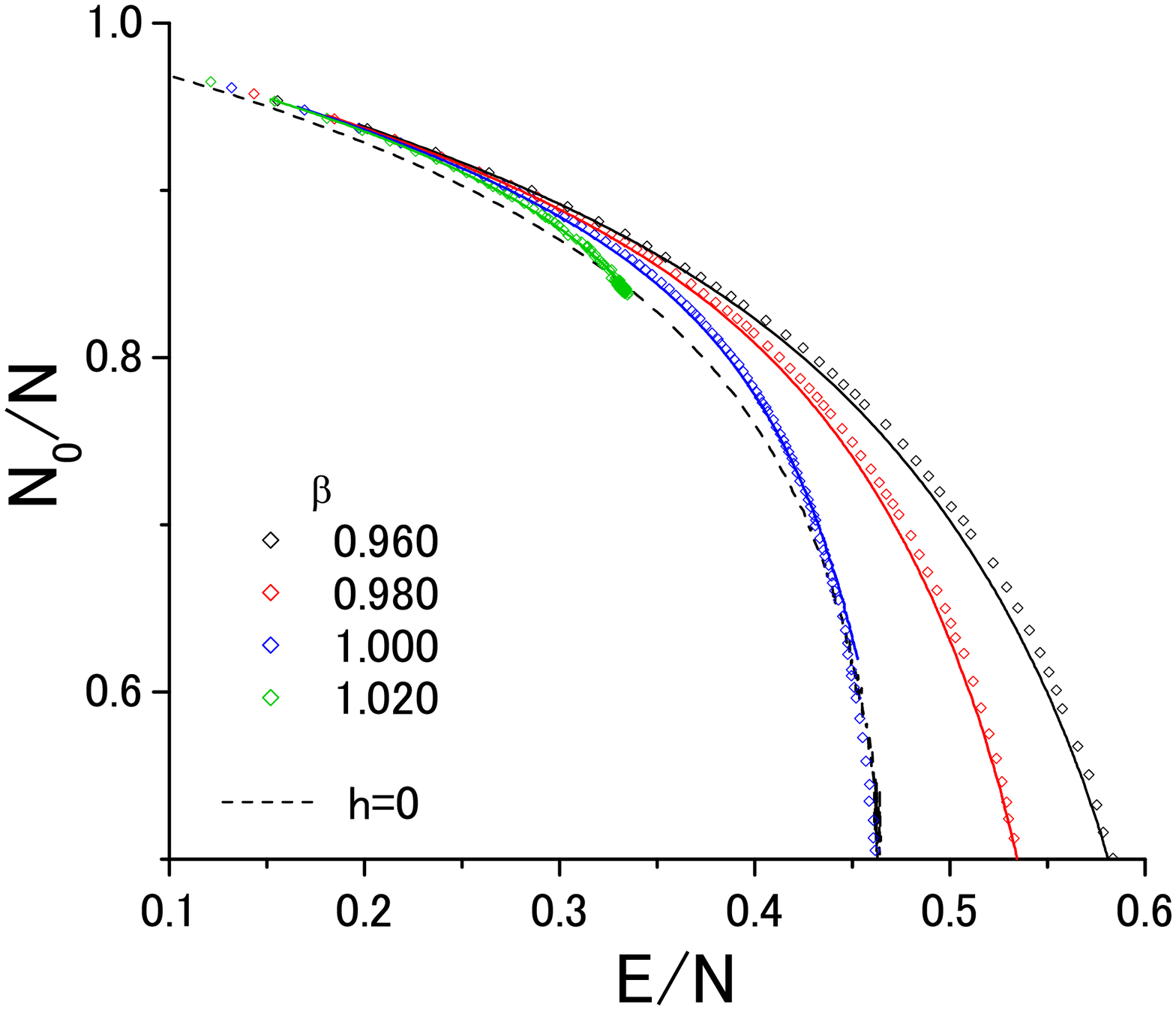}
\\ \hspace{-6.8cm} {\bf{\large (b)}}\\ \vspace{-14.1pt}
\includegraphics[trim=20 40 140 20,scale=0.32,clip]{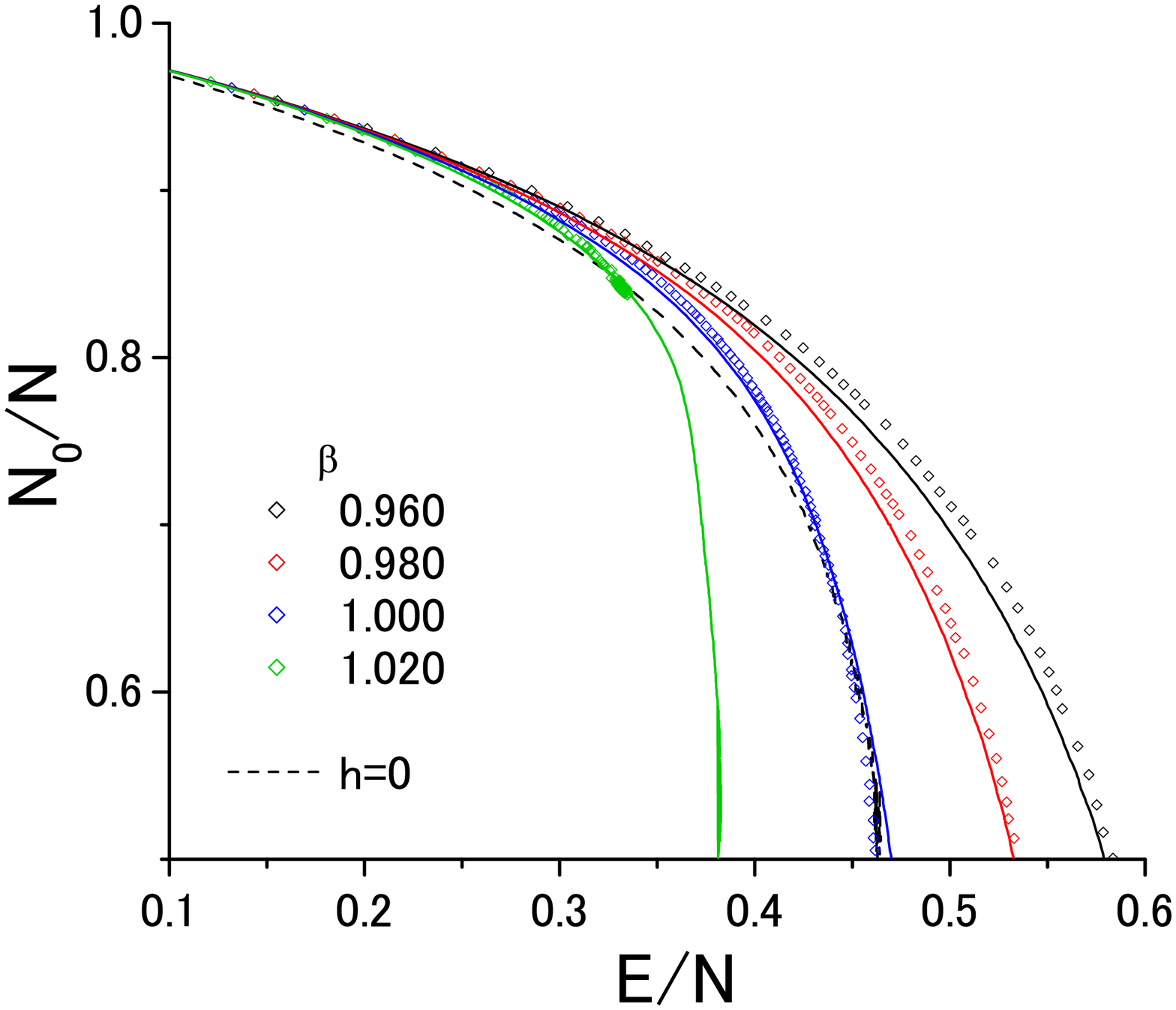}
\end{center}
\vspace{-5mm}
\caption{\label{fig:trajectory}
(color online) 
(a) Trajectories for the system $L=128$.
The results obtained by the integration of the master equation are denoted by lines
and those obtained by the KMC method are denoted by symbols.
The results of the two-variable description reproduce the true dynamics significantly well.
The contour line $h(E,N_0) = 0$ (dashed line) is also shown. 
(b) The contour line of $\beta(\AA)$. 
}
\end{figure}

Figure~\ref{fig:xy} shows the flow diagram in the $(E \times N_0)$-space at $\beta=1.010>\beta_c$. 
The flow line is drawn by trajectories 
that are obtained by integrating $\VV$ for various initial conditions. 
The trajectories rapidly join the unique one flowing into the fixed point. 
Because of the rotation-free velocity field, 
the nonequilibrium free energy Eq.~(\ref{eq:F_neq}) can be defined, 
and it differs from the equilibrium one.
In Fig.~\ref{fig:trajectory}(a), trajectories obtained by the master equation 
and KMC are plotted and show good agreement for each temperature.

The line $h(E,N_0) \equiv-\del \ln g(E,N_0)/\del N_0 = 0$ is shown in Fig.~\ref{fig:xy}. 
This line represents the state equation without the magnetic field. 
[The left below region of the line is the coexisting phase 
of domains ordered to states 0, 1, and 2. 
To be precise, $h(E,N_0)$ also equals zero in this region in the thermodynamic limit.]
Each point on this line is related to the most probable state 
in the fixed $E$ ensemble, i.e., the microcanonical ensemble, without the magnetic field. 
Therefore, in the one-variable description, 
the transient state is confined to this equilibrium line independently of $\beta$. 
On the other hand, the trajectory of the disordering process 
in the two-variable description runs above and to the right of the equilibrium line 
and stops when touching the line. 
By adding the second variable $N_0$, 
the system can take a path away from the equilibrium line depending on $\beta$. 
This is the reason for the drastic improvement in accuracy from the one-variable description.

\subsection{Comparison with the free energy valley}

While the contour line $h(E,N_0) = 0$ is related to the equilibrium line 
in the fixed-$E$ ensemble for $h=0$, 
$\beta(E,N_0) = \beta_\mrm{ext}$ is related to the equilibrium line 
for the fixed-$N_0$ ensemble for a given $\beta_\mrm{ext}$. 
The latter gives a trajectory of the quasi-equilibrium dynamics 
when the relaxation of the magnetization is much slower than that of the energy. 
Figure~\ref{fig:trajectory}(b) shows the contour lines of $\beta$. 
The contour lines are similar to the trajectory 
of the isothermal relaxation obtained by the KMC method 
although the agreement is  slightly less than that of the master equation results 
[Fig.~\ref{fig:trajectory}(a)] far from equilibrium. 
The quasi-equilibrium dynamics cannot reproduce the rapid change in $N_0$, 
but that has an almost negligible effect in the present case. 
This is a useful approximation for the trajectory; 
note, however, that the contour lines of the temperature provide 
no information on the time evolution.


\section{Summary and Discussion}
\label{sec:discussion}

In this paper, we propose a method to analyze the non-stationary nonequilibrium state 
by using a master equation for extensive (macroscopic) variables 
and show some applications of the method. 
The method uses the quantity $\mu(\AA; \DAA)$ calculated by equilibrium ensemble for fixed EVs, 
which means an equal probability is assumed 
for microscopic states in nonequilibrium. 

For the all-to-all coupling model, the state can be identified accurately 
only by  the single variable $N_0$ since the system lacks a spatial structure. 
Thus, there is no difference between equilibrium and nonequilibrium states in this case, 
and the one-variable description gives rigorous dynamics. 
For the two-dimensional 3-state Potts model, while the one-variable description  
is far from adequate to resemble the true dynamics, the two-variable description is sufficient. 
This improvement can be attributed to the fact that 
we have a unique trajectory for any external conditions $\beta$ 
in the one-dimensional $A$-space corresponding to $E$; 
this trajectory is equivalent to the equilibrium line in two dimensions 
(this is the reason why strange inflection points, 
which corresponds to the singularity at $\beta=\beta_c$, 
are observed for off-critical temperatures as in Fig.~\ref{fig:1d_V-E}(a).)
It has been suggested that transient states in finite-dimensional systems 
have distinctly different kinds of spatial fluctuations 
depending on temperature even for the same internal energy. 
Specifically, the trajectory runs in the region where $h(E,N_0)>0$. 
For the same $E$, $N_0$ on the trajectory is larger for higher temperatures. 
This is because the minority spins with number $(N - N_0)$ 
have a lesser tendency to aggregate at higher temperatures 
and thus acquire $E$ with smaller $(N-N_0)$ from the surface energy.



To investigate equilibrium critical properties 
from nonequilibrium relaxation at the critical point, 
the so-called nonequilibrium relaxation analysis \cite{Ozeki07} can be used.
In the relaxation, a nonequilibrium correlation length 
grows with time in a power law, 
which is similar to the growth of the equilibrium correlation length 
when approaching the critical point by tuning an external parameter. 
The critical exponents are estimated using this similarity \cite{Ito00}. 
The two paths on which the growth of the nonequilibrium and equilibrium correlation length 
are observed correspond to the trajectory of $\beta=\beta_c$ 
and the equilibrium line $h(E,N_0)=0$, respectively.
Both paths approach the critical point $(E, N_0)=N(e_c, 1/3)$, 
where $e_c \equiv 1-1/\sqrt{3}$, and the paths are expressed 
as $(N_0/N - 1/3)^{1/\beta} = C (e_c - E/N)^{1/(1-\alpha)}$. 
However, the proportionality constant $C$ is different between the two, 
which makes the two paths macroscopically distinguishable. 
Interestingly, while the relaxation dynamics along them are different, 
the critical exponents, such as $\alpha$ and $\beta$, are the same.

For the two-dimensional 3-state Potts model, 
the second variable, the order parameter, has special importance 
because its relaxation is much slower than that of the energy. 
Thus, it has more impact on the total dynamics. 
Therefore, the quasi-equilibrium dynamics with respect to the order parameter 
is a good approximation.
Indeed, the contour line of the temperature, i.e., 
the array of equilibrium points for various temperatures, 
gives a good analog of the relaxation trajectory 
it does not provide information on the time evolution. 
One would think that the one-variable description with $N_0$ would give good results 
comparable to those of the two-variable description with $E$ and $N_0$. 
However, in our formulation, the EV conjugate to the non-zero intensive variable 
must be employed to determine $w(\bb; \DAA)$.

It is possible to improve the accuracy of our method by employing more DOFs; 
however, the difficulty is to find appropriate variables independent of each other. 
Important criteria are the slowness of the relaxation 
and the capture of the spatial structure. 
One good candidate would be the structure factor, 
$S_0(\qq) = |\sum_i \delta_{\sigma_i 0} e^{i \qq \cdot \rr_i}|^2/N$.
The quantity $N_0$ is equivalent to $\sqrt{N S_0(\mbf{0})}$. 
By adding $S_0$ from small $\qq$ to larger ones, 
a finer structure of the space would be included stepwise. 
However, in practice, the addition of DOFs is very difficult 
in the implementation adopted here 
because the dimensionality of the $A$-space increases. 
Depending on the problem, we might not need information 
of the wide range of the $A$-space. 
In addition, the calculation of $\mu$ is suitable for trivial parallel computation.

One relaxation dynamics of a given intensive variable can be generated  
with a reasonably lower computational cost by KMC method than by the present method. 
Our method takes significant computational time to estimate $\mu(\AA;\DAA)$. 
Once this is obtained, however, one can easily generate time evolutions 
for various $\bb$. 
This is similar to the efficient calculation method of equilibrium expectation values 
by using the reweighting method with extended ensembles \cite{Landau04}. 
This aspect is very useful when calculating the fluctuation of quantities, 
which requires many samples. 
However, our main aim is not to develop an efficient way to yield time evolution 
but to find a way to understand the system far from equilibrium 
in the framework of equilibrium statistics. 
The basic concept proposed here is not only for numerical analysis but for analytical treatment. 
It is possible to develop an approach to nonequilibrium dynamics 
by utilizing the attained knowledge of equilibrium physics. 
The success in the present demonstration encourages us to continue in this direction.

We have found that the present method works well for critical relaxation dynamics. 
The next interesting application is the dynamics taking first-order transitions \cite{Rikvold94}, 
such as liquid-vapor transition and protein folding \cite{Frauenfelder91}.
It is often explained that the nucleation dynamics is governed by the free-energy barrier  
determined by the competition between bulk and surface free energies. 
However, for a finite-dimensional system with short-range interaction, 
this picture is not applicable to the macroscopic free energy, 
which is a convex function of EVs \cite{Nogawa11b}. 
Therefore, the addition of DOFs is essential for nucleation dynamics.

The basic idea of the present method can be applied to a non-stationary state 
of an open system with certain driving forces; 
a non-stationary state is described by a nonequilibrium stationary state.
By replacing the word ``equilibrium'' with ``stationary'', 
we obtain a parallel framework for transient dynamics toward stationary states.

\vspace{.20cm}
\begin{center}
{\bf ACKNOWLEDGMENTS}
\end{center}

\vspace{-.20cm}

This work was partly supported by Award No. KUK-I1-005-04 
presented by King Abdullah University of Science and Technology (KAUST).



%

\end{document}